\documentclass[twoside]{article}
\usepackage{fleqn,espcrc2}
\usepackage{graphicx}
\usepackage{here}

\newcommand{\AmS}{{\protect\the\textfont2
    A\kern-.1667em\lower.5ex\hbox{M}\kern-.125emS}}
\def\beq{\begin{equation}}
\def\eeq{\end{equation}}
\def\bea{\begin{eqnarray}}
\def\eea{\end{eqnarray}}
\def\bq{\begin{quote}}
\def\eq{\end{quote}}

\def\la{\langle}
\def\ra{\rangle}

\def\ba{\begin{array}}
\def\ea{\end{array}}

\def\gg2{ \la\alpha_s G^2 \ra}
\def\gg3{g^3f_{abc}\la G^aG^bG^c \ra}

\usepackage{epsfig}

\usepackage{amssymb}

\title{\bf{\boldmath
{\Large Yang-Mills Propagators and QCD }\thanks{Talk given at QCD 08 International Conference (Montpellier, 7-12th July 2008)}}}

\author{Marco Frasca \\
Via Erasmo Gattamelata, 3 \\
00176 Roma (Italia)}

\begin{document}

\begin{abstract}
\noindent
We present a strong coupling expansion that permits to develop analysis of quantum field theory in the infrared limit.
Application to a quartic massless scalar field gives a massive spectrum and the propagator in this regime. We extend the
approach to a pure Yang-Mills theory obtaining analogous results. The gluon propagator is compared satisfactorily
with lattice results and similarly for the spectrum. Comparison with experimental low energy spectrum of QCD
supports the view that $\sigma$ resonance is indeed a glueball. The gluon propagator we obtained is finally used to
formulate a low energy Lagrangian for QCD that reduces to a Nambu-Jona-Lasinio model with all the parameters
fixed by those of the full theory.
\end{abstract}

\maketitle







One of the main difficulties people meets to cope with low energy limit of a quantum field theory is the missing of
a perturbation techniques that could permit to extract analytical manageable results to compare with experiment.
In the fall of seventies, Bender's group proposed a possible method to solve this impasse \cite{be1,be2,be3,be4}. 
Bender's group approach can be understood with a simple quartic oscillator as (here and following $\hbar=c=1$)
\begin{equation}
    G(t_2-t_1)=\int [dq(t)] e^{i\int_{t_1}^{t_2}dt
    \left[\frac{1}{2}\dot q(t)^2-\frac{\lambda}{4}q(t)^4\right]}.
\end{equation}
When the limit $\lambda\rightarrow\infty$ is taken one simply neglects the kinematic term assuming it as a higher
order correction leaving us with 
\begin{equation}
    G(t_2-t_1)\approx\int [dq(t)] e^{-i\int_{t_1}^{t_2}dt\frac{\lambda}{4}q(t)^4},
\end{equation}
a highly singular expression that needs to be regularized with a proper cut-off. 
This in turn implies that computations produce series generally difficult to resum due to the very singular dependence on the cut-off, 
and the method proves to be not effective in the computation of propagators and spectra. The reason why Bender's group method does not work
properly is due to the completely omitted dynamics that should be permitted instead. This can be accomplished
if we do a time rescaling as $t\rightarrow\sqrt{\lambda}t=\tau$ and we are left with the expression
\begin{equation}
    G(\tau_2-\tau_1)=\int [dq(\tau)] e^{i\sqrt{\lambda}\int_{\tau_1}^{\tau_2}d\tau
    \left[\frac{1}{2}\dot q(\tau)^2-\frac{1}{4}q(\tau)^4\right]}
\end{equation}
that is perfectly regular being just the semiclassical limit when $\lambda\rightarrow\infty$. 
In this case the spectrum is computed straightforwardly and, accounting for all higher order
corrections, one has the exact result \cite{be5}. Starting from the Schr\"odinger equation
\begin{equation}
    -\frac{\hbar^2}{2m}\frac{\partial^2\psi}{\partial x^2}+\frac{\lambda}{4}x^4\psi=
	i\hbar\frac{\partial\psi}{\partial t},
\end{equation}
rescaling time as $\tau=\sqrt{\lambda}t$ one has
\begin{equation}
    -\frac{1}{\lambda}\frac{\hbar^2}{2m}\frac{\partial^2\psi}{\partial x^2}+\frac{1}{4}x^4\psi=
    i\hbar\frac{\partial\psi}{\partial\tau}
\end{equation}
and finally taking an expansion like $\psi = \psi^{(0)}+\frac{1}{\lambda}\psi^{(1)}+
\frac{1}{\lambda^2}\psi^{(2)}+O\left(\frac{1}{\lambda^3}\right)$ for the wave function,
the above method recovers a well-known semiclassical expansion, 
\begin{eqnarray}
   \psi &=& e^{-\frac{i}{\hbar}\frac{x^4}{4}\tau}\left\{\phi(x)-\frac{\hbar^2}{2m}\right.
   \left[\frac{\partial^2\phi(x)}{\partial x^2} \right. \\
   & &-2\frac{i}{\hbar}x^3\tau \frac{\partial\phi(x)}{\partial x}
   \left.\left.-\left(3\frac{i}{\hbar}x^2\tau+\frac{1}{\hbar^2}x^6\tau^2\right)\phi(x)\right]+\ldots\right\} \nonumber
\end{eqnarray}
that is the Wigner-Kirkwood expansion with the spectrum given by th usual WKB series \cite{fra1}. But 
Wigner-Kirkwood expansion is a {\sl gradient expansion} and this is the key result:
{\sl A strongly coupled quantum system behaves semiclassically and its behavior is 
described by a Wigner-Kirkwood gradient expansion.} A typical application for
this fundamental result could be measurement theory when a measuring apparatus is strongly
coupled to a quantum system. But the method is quite general to be applicable to
whatever differential equation.

We would like to extend the above approach to quantum field theory. For our aims, we
consider a massless quartic scalar field theory with a generating functional
\begin{equation}
    Z[j] = \int[d\phi]e^{i\int d^4x\left[\frac{1}{2}(\partial\phi)^2-\frac{\lambda}{4}\phi^4+j\phi\right]}
\end{equation}
and rescale time as shown above for the quartic oscillator to obtain
\begin{equation}
     Z[j] = \int[d\phi]e^{i\sqrt{\lambda}\int d^4x\left[\frac{1}{2}(\dot\phi)^2-\frac{1}{4}\phi^4+j\phi\right]}
                     e^{-\frac{i}{\sqrt{\lambda}}\int d^4x\frac{1}{2}(\nabla\phi)^2}
\end{equation}
that in the limit $\lambda\rightarrow\infty$ permits us to recover again the semiclassical limit and a
gradient expansion \cite{fra2}. But in this case we are able to solve the equation for the propagator
\begin{equation}
      \ddot G(t_1-t_2)+\lambda G(t_1-t_2)^3=\mu^2\delta(t_1-t_2)
\end{equation}
that is
\begin{equation}
\label{eq:prop}
    G(t_1-t_2) = \theta(t_1-t_2)\mu\left(\frac{2}{\lambda}\right)^{\frac{1}{4}}
    {\rm sn}\left[\left(\frac{\lambda}{2}\right)^{\frac{1}{4}}\mu t\right]
\end{equation}
being sn a Jacobi elliptic function and $\mu$ an arbitrary integration constant. Now, using
a small time approximation we can take the following definition for the field $\phi$ \cite{fra3,fra4}
\begin{equation}
    \phi(t) \approx \int dt'G(t-t')j(t')
\end{equation}
we can restate the above generating functional for the scalar field theory in a Gaussian form
\begin{equation}
    Z[j]\approx e^{\frac{i}{2}\int d^4y_1d^4y_2\frac{\delta}{\delta j(y_1)}(-\nabla^2)\delta^D(y_1-y_2)
    \frac{\delta}{\delta j(y_2)}} Z_0[j]
\end{equation}
being
\begin{equation}
    Z_0[j]=e^{\frac{i}{2}\int d^4x_1d^4x_2j(x_1)\Delta(x_1-x_2)j(x_2)}
\end{equation}
with $\Delta(x_1-x_2)=\delta^3(x_1-x_2)[\theta(t_1-t_2)G(t_1-t_2)+\theta(t_2-t_1)G(t_2-t_1)]$. Having obtained
the propagator in the strong coupling limit we can extract the spectrum of the theory in the infrared limit. This
can be done using eq.(\ref{eq:prop}) and the relation for the sn Jacobi function \cite{gr}
\begin{equation}
    {\rm sn}(u,i)=\frac{2\pi}{K(i)}\sum_{n=0}^\infty
      \frac{(-1)^ne^{-(n+\frac{1}{2})\pi}}{1+e^{-(2n+1)\pi}}
      \sin\left[(2n+1)\frac{\pi u}{2K(i)}\right]
\end{equation}
being $K(i)=\int_0^{\frac{\pi}{2}}\frac{d\theta}{\sqrt{1+\sin^2\theta}}\approx 1.3111028777$.
This gives in our case for the Feynman propagator
\begin{equation}
    \Delta(\omega)=\sum_{n=0}^\infty
    (2n+1)\frac{\pi^2}{K^2(i)}\frac{(-1)^{n+1}e^{-(n+\frac{1}{2})\pi}}{1+e^{-(2n+1)\pi}}
    \frac{1}{\omega^2-\omega_n^2+i\epsilon}
\end{equation}
with the harmonic oscillator spectrum
\begin{equation}
    \omega_n = \left(n+\frac{1}{2}\right)\frac{\pi}{K(i)}
    \left(\frac{\lambda}{2}\right)^{\frac{1}{4}}\mu.
\end{equation}
We see immediately that the propagator is in agreement with K\"allen-Lehman representation. This is 
a key observation if we want to interpret the poles of the propagator as physical states
even if the propagator is gauge dependent. But we also see that
this propagator satisfies the Callan-Symanzik equation
\begin{equation}
    \mu\frac{\partial G(t)}{\partial\mu}+
   \beta(\lambda)\frac{\partial G(t)}{\partial\lambda}+2\gamma G(t)=0
\end{equation}
with a beta function $\beta = -4\lambda$ and $\gamma = -1$. This means that the
theory is trivial in the infrared limit as the coupling goes to zero as the
fourth power of momentum \cite{fra5}.

The next step is to apply the above approach to a pure Yang-Mills theory in order to analyze the behavior
in the low energy limit. But we have to cope here with a serious difficulty. In the eighties it was
proved by Matinyan, Savvidy, et al. \cite{sav1,sav2,sav3} that the classical equations of motion of
Yang-Mills theory are generally chaotic and so, completely useless to built a quantum field theory.
The way out is the following mapping theorem \cite{fra6}:

{\bf MAPPING THEOREM:} {\sl An extremum of the action
\begin{equation}
    S = \int d^4x\left[\frac{1}{2}(\partial\phi)^2-\frac{\lambda}{4}\phi^4\right]
\end{equation}
is also an extremum of the SU(N) Yang-Mills Lagrangian when 
we properly choose $A_\mu^a$ with some components being zero and all others being equal, and
$\lambda=Ng^2$, being $g$ the coupling constant of the Yang-Mills field.}

\begin{figure}[tbp]
\begin{center}
\includegraphics[width=250pt]{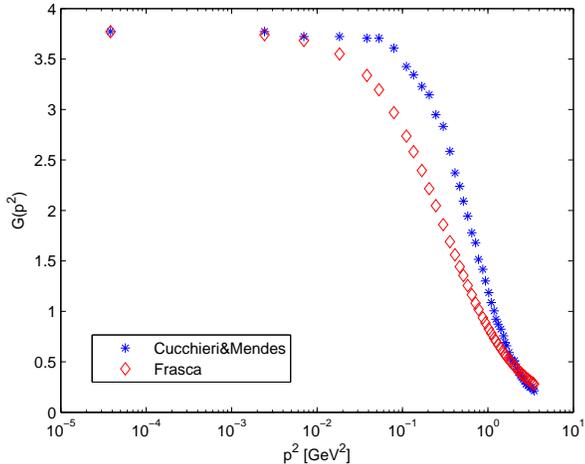}
\caption{\label{fig:fig1} Fit to lattice data of propagator in eq. (\ref{eq:ym}).}
\end{center}
\end{figure}

This theorem holds exactly at classical level but at a quantum level can be maintained only in
the infrared limit as in the ultraviolet case quantum fluctuations spoil the mapping
producing asymptotic freedom, a property missing for the scalar field \cite{ps}. Then,
using mapping theorem we can write down immediately the gluon propagator in the Landau gauge as
\begin{eqnarray}
\label{eq:ym}
     D_{\mu\nu}^{ab}(p)&=&\delta^{ab}\left(\eta_{\mu\nu}-\frac{p_\mu p_\nu}{p^2}\right)\sum_{n=0}^\infty
    (2n+1)\frac{\pi^2}{K^2(i)} \times \\
    & &\frac{(-1)^{n+1}e^{-(n+\frac{1}{2})\pi}}{1+e^{-(2n+1)\pi}}\frac{1}{p^2-m_n^2+i\epsilon} \nonumber
\end{eqnarray}
where we have also introduced Latin indexes for color degree of freedom and 
\begin{equation}
\label{eq:mn}
    m_n = \left(n+\frac{1}{2}\right)\frac{\pi}{K(i)}
    \left(\frac{Ng^2}{2}\right)^{\frac{1}{4}}\Lambda.
\end{equation}
being $\Lambda$ the integration constant.
Mapping theorem tells us
also that the ghost field decouples from the gluon field and must behave like a free particle
with a propagator $G(p)=1/(p^2+i\epsilon)$. The running coupling must go to zero like
the fourth power of momentum being this in agreement with lattice computations as shown
by Boucaud et al. \cite{bou} but as also seen from experiments as shown by Prosperi et al. \cite{pro1,pro2,pro3}.
These latter results on the running coupling give a strong clue that Yang-Mills theory is trivial
as happens for the scalar field theory. Lattice computations give also results on the propagators \cite{ilg2,cuc,ste}.
These show that the ghost is very near the free particle case while the gluon propagator reaches a finite
value at zero momentum. So, we compare our propagator (\ref{eq:ym}) with lattice computations taking the
one with the largest volume (27fm)$^4$ \cite{cuc}. For this to work we need to go to Euclidean momenta and
fix $m_0$ in eq.(\ref{eq:mn}). The result is presented in Fig. \ref{fig:fig1} and is
really satisfactory. 

The comparison is done in our case with $m_0=\frac{\pi}{2K(i)}\left(Ng^2/2\right)^{\frac{1}{4}}\Lambda\approx 547\ MeV$
corresponding to a string tension $\sqrt{\sigma}=\left(Ng^2/2\right)^{\frac{1}{4}}\Lambda\approx 456\ MeV$. This result
is striking indeed as we observe a mass for the ground state very near that of the $\sigma$ resonance and a string
tension very near that generally obtained from experiments $410-440\pm 20 MeV$. So, on the lattice the scenario
is clearly settled. Numerical solution of Dyson-Schwinger equations also confirm it \cite{an}. 

For the spectrum the situation is satisfactory as well. Comparing with lattice computations by Teper et al. \cite{tep1,tep2},
we have shown in \cite{fra7} that one gets Tab.\ref{tab:0++} and \ref{tab:2++}. Theoretical numbers compared in the tables
are computed through the relation $\frac{m_n}{m_0}=(2n+1)\frac{\pi}{2K(i)}$ that defines a set of ``golden numbers'' that
are those measured in lattice computations of spectra.

\begin{table}[tbp]
\begin{center}
\begin{tabular}{|c|c|c|c|} \hline\hline
Excitation & Lattice & Theoretical & Error \\ \hline
$\sigma$   & -       & 1.198140235 & - \\ \hline 
0$^{++}$   & 3.55(7) & 3.594420705 & 1\% \\ \hline
0$^{++*}$  & 5.69(10)& 5.990701175 & 5\% \\ \hline\hline
\end{tabular}
\caption{\label{tab:0++} Comparison for the 0$^{++}$ glueball}
\end{center}
\end{table}

\begin{table}[tbp]
\begin{center}
\begin{tabular}{|c|c|c|c|} \hline\hline
Excitation & Lattice & Theoretical & Error \\ \hline
$\sigma^*$ & -       & 2.396280470 & - \\ \hline 
2$^{++}$   & 4.78(9) & 4.792560940 & 0.2\% \\ \hline
2$^{++*}$  & -       & 7.188841410 & - \\ \hline\hline
\end{tabular}
\caption{\label{tab:2++} Comparison for the 2$^{++}$ glueball}
\end{center}
\end{table}

We cannot say the same with the data by C. Morningstar et al. \cite{mor} but these authors use anisotropic lattices. 
The $\sigma$ state appears on lattice computations of the gluon propagator but not on lattice computations of spectra. It
should be said that presently the volumes used in the former computations are far larger than for the latter case.
Our view is that an effort should be done in this direction in order to get a clear understanding of Yang-Mills
theory on the lattice by increasing the volumes for spectrum computation. 

A needed step is to understand current phenomenology of the low energy spectrum of QCD with the above results. Indeed,
the ground state of a pure Yang-Mills theory can be identified with the observed resonance $\sigma$ or $f_0(600)$. This means 
that Yang-Mills theory has a higher ground state than full QCD as quarks lower it to the pion mass. Indeed,
 $f_0(600)$ or $\sigma$ seems to have a large gluonic content due to the very small width
 for $\gamma\gamma$ decay that can be due to $\pi\pi$ rescattering \cite{nar1,nar2}
 and as also seen through $\chi$PT \cite{pel1,pel2}. We have shown that
our computed mass is very near the one obtained from experiments \cite{bug}. If we take for the string tension 440 MeV as is customary in lattice computations we get $m_\sigma=527 MeV$ and, as seen above, lattice computations for the gluon propagator give about 547 MeV. Mixing with quarks could reduce this value. A more conservative choice of 410 MeV can take both $\sigma$ and $\sigma^*$, the latter eventually being $f_0(980)$, to
agree with measured data giving $m_\sigma\approx 491\ MeV$ and $m_{\sigma^*}\approx 982\ MeV$. This view is shared in recent analysis \cite{nar1,nar2}. But while these resonances are seen both experimentally and theoretically and, in indirect way through lattice computations for the propagator, a direct evidence on lattice computations for spectra appears difficult to achieve presently \cite{mcn}.

As pointed out in \cite{tg}, once gluon propagator is known one can obtain a low energy model directly from QCD. 
We derived this model in \cite{fra8}. So, taking QCD action as
\begin{eqnarray}
    S &=& \int d^4x\left[\sum_q\bar q\left(i\gamma\cdot\partial - g\frac{\lambda^a}{2}\gamma\cdot A^a - m_q\right)q\right. \\
   & &\left.-\frac{1}{4}G^a_{\mu\nu}G^{a\mu\nu}-\frac{1}{2\alpha}(\partial\cdot A)^2\right] \nonumber
\end{eqnarray}
being $G^a_{\mu\nu}=\partial_\mu A^a_\nu-\partial_\nu A^a_\mu+gf^{abc}A^b_\mu A^c_\nu$,
we use the small time approximation seen for the quartic scalar and Yang-Mills fields \cite{fra3,fra4} with
the above computed propagator writing
\begin{equation}
    A^a_\mu(x)\approx g\int d^4y  D_{\mu\nu}^{ab}(x-y)\sum_q\bar q(y)\frac{\lambda^a}{2}\gamma^\nu q(y)
\end{equation}
giving
\begin{eqnarray}
   S &\approx& \int d^4x\left[\sum_q \bar q(x)(i\gamma\cdot\partial-m_q)q(x) \right.\\
   & &-\frac{1}{2}g^2\sum_{q,q'}\bar q(x)\frac{\lambda^a}{2}\gamma^\mu q(x)\times \nonumber \\
   & &\left.\int d^4y D_{\mu\nu}^{ab}(x-y)\bar q'(y)\frac{\lambda^b}{2}\gamma^\nu q'(y)\right]. \nonumber
\end{eqnarray}
In the infrared limit $p\ll m_0$, being $m_0=(Ng^2/2)^{\frac{1}{4}}\pi/2K(i)\Lambda=m_\sigma\approx 491 MeV$ 
that can be taken as the NJL cut-off $\Lambda_{NJL}$, one gets
\begin{eqnarray}
   S &\approx& \int d^4x\left[\sum_q \bar q(x)(i\gamma\cdot\partial-m_q)q(x)\right.\\ 
   & &-\frac{1}{2}g^2\sum_{q,q'}\bar q(x)\frac{\lambda^a}{2}\gamma^\mu q(x)\times \nonumber \\
   & & \left.\int d^4yG(x-y)\bar q'(y)\frac{\lambda^a}{2}\gamma_\mu q'(y)\right]. \nonumber
\end{eqnarray}
being $G(x-y)\approx 3.761402959\cdot\frac{1}{\sigma}\delta^4(x-y)$ (Fermi approximation). This gives in the end
\begin{eqnarray}
   S &\approx& \int d^4x\left[\sum_q \bar q(x)(i\gamma\cdot\partial-m_q)q(x)\right. \\
   & &-\left.\frac{1}{2}G_{NJL}\sum_{q,q'}\bar q(x)\frac{\lambda^a}{2}\gamma^\mu q(x)
   \bar q'(x)\frac{\lambda^a}{2}\gamma_\mu q'(x)\right] \nonumber
\end{eqnarray}
being $G_{NJL}\approx 3.761402959\cdot\frac{g^2}{\sigma}$ that links QCD theory parameters and the coupling of the NJL model.
The point opened up by the introduction of quarks is how to recover gluonic contributions to QCD spectrum. This possibility
can be achieved through bosonization techniques \cite{ebe}.

So, a consistent formulation of quantum field theory in the infrared exists. We have seen that a satisfactory
agreement can be obtained in this way with lattice computations both for the propagators and the spectrum
for Yang-Mills theory. But while for the propagators the agreement is really good, for the spectrum some
more effort on lattice computations seems needed to obtain a satisfactory comprehension of all the matter. 
From the experimental side the situation is quite good granting an understanding of where glueballs should 
lie in the observed spectrum. Agreement is seen between different theoretical approaches in this case while
some more work is needed to clarify the situation. It should be said that if further elements will confirm
this scenario, surely a lot of unexpected views about strong interactions will force new understanding for all
quantum field theory.




\end{document}